\def\year{2021}\relax
\documentclass[letterpaper]{article} % DO NOT CHANGE THIS
\usepackage{aaai21}  % DO NOT CHANGE THIS
\usepackage{times}  % DO NOT CHANGE THIS
\usepackage{helvet} % DO NOT CHANGE THIS
\usepackage{courier}  % DO NOT CHANGE THIS
\usepackage[hyphens]{url}  % DO NOT CHANGE THIS
\usepackage{graphicx} % DO NOT CHANGE THIS
\usepackage{natbib}  % DO NOT CHANGE THIS AND DO NOT ADD ANY OPTIONS TO IT
\usepackage{caption} % DO NOT CHANGE THIS AND DO NOT ADD ANY OPTIONS TO IT
\title{The Definition-Context-Purpose Paradigm and Other Insights from Industry Professionals About the Definition of a Quest}
\author{
    Kristen K. Yu, 
    Matthew Guzdial and
    Nathan R. Sturtevant \\
}
\affiliations{
    Department of Computing Science, University of Alberta\\
    Alberta Machine Intelligence Institute (AMII)\\
    kkyu@ualberta.ca, guzdial@ualberta.ca, nathanst@ualberta.ca
}
\begin{document}

\maketitle

\begin{abstract}
Among academic communities there is no single agreed upon definition of a quest. The industry perspective on this topic is also largely unknown. Thus, the purpose of this paper is to gain an understanding of the definition of a quest from industry professionals to better inform the academic community. We interviewed fifteen game developers with experience designing or implementing quests or narratives, and process the interviews using thematic analysis to identify trends. We identified a variety of personal developer definitions. However, we also discovered several themes that may inform future academic work. We introduce the definition-context-purpose paradigm as a synthesis of these trends: elements of a quest, purpose of a quest, and context of a quest. Finally, we discuss the developer’s reaction to a recently proposed quest definition as part of a push towards a general quest definition. 

\end{abstract}

%IN THE INTRO: Prior work has been trying to "reverse engineer" what a ques tis,
%this is the first instance of "deductive" quest theory generation.

%Results:

%Make a sequence of point:
%1) Quest definitions are all over the place. This is indicative of a lack of %thought. "I've never thought about it before", changing/modifying their definition %during the thing.
%2) Then we can say, despite that, there is overlap. So when we ask them what a %quest definition is.
%3) Go into more detail-> how do they overlap, and what are the ways that they %overlap,
%4) The way that designers think aobut quests that are unique -> Split things %between definition and purpose.
%5) My quest definition, ending with the response to my definition
%-> There is some notion that we are getting towards this definition that has this %broad appeal.
%

\section{Introduction}
 
Video game quests are an area of interest for many academic research fields. Quests are often procedurally generated \cite{doran_prototype_2011}, are fundamental to interactive narratives \cite{li_offline_2010}, and are used to help provide more personalized experiences \cite{vanhatupa_guidelines_2011}. Research projects in these areas typically do not offer a formal definition of a quest, and assume that it common knowledge in order to focus on other problems instead. In our previous research we showed there is no consensus on the technical definition of a quest \cite{yu_what_2020}, and each definition has limited applicability to other research. Industry professionals are a natural source of knowledge for a quest definition, so in this paper we interview quest and narrative developers. 

%Video game quest theory was first introduced by \citeauthor{aarseth_hunt_2005} \cite{aarseth_hunt_2005}. Early work on the definition of a quest focused on the distinctions between quests in video games and in novels \cite{aarseth_quest_2004, tronstad_semiotic_2001}. Some researchers brought ideas from other popular media such as literary work or Dungeons and Dragons \cite{DnD} to help create their definitions of a quest \cite{howard_quests_2008, tosca_quest_2003}. While other areas of quest theory focused on deriving design patterns from games \cite{doran_towards_2010, smith_situating_2011-1, karlsen_quests_2008}. 
%This proved to be a popular technique to further quest research as the field continued.

%which proved to be a popular way to further quest research as the field continued.

%In addition to analyzing games, some researchers chose to bring ideas from other popular media such as literary works or Dungeons and Dragons 

Research on video game quest theory appears to have lost popularity in recent years. Modern quest theory is typically researched in service of other goals, and focuses on other areas rather than the quest definition \cite{alexander_deriving_2017, mourato_challenge_2013, machado_towards_2016, sullivan_rules_2010, dahlskog_patterns_2015, lee_effects_2009}. Since there is interest in these other areas, it is worth re-examining the definition of a quest. Many research areas rely on quest definitions, so they each create a quest definition independently. Oftentimes, these definitions are not technically rigorous, and assume words in the definition are common knowledge. If the academic community had a general quest definition to reference we could increase collaboration, more easily compare quest related work, and support a methodical exploration of quests.

%For example, quests were analyzed for mechanic or structural similarities in order to more effectively procedurally generate quests . Quest theory has also been used to help design levels \cite{dahlskog_patterns_2015} or model player knowledge \cite{lee_effects_2009}. 

%The early work on quest theory, and many later works done in service for a particular goal, all share similar problems in fully defining a quest. The most common limitation is the context in which the quest definition is derived from. Researchers mostly analyze quests in RPGs or MMORPGs, which limit the applicability of the quest definition to those genres. In the past, it may have been appropriate to limit video game quest definitions to specific games or genres, but in the past ten years it has been increasingly common for games to borrow elements from other genres. One of the more popular additions is the inclusion of quests in non-RPG games, such as the Rocket Pass in the sports game Rocket League \cite{RocketLeague}. This illustrates the need to update the definition of a quest to maintain relevance for modern games.  If the academic community has a single, generalized quest definition to reference, quests can be more easily applied across research areas such as quest generation, interactive narrative, and player modeling. 

%Instead, there could be a vocal minority of developers who are dominating the public discourse. 

Research on the definition of a quest would benefit from the perspective of developers. Identifying design patterns are a common way for researchers to incorporate the industry perspective on quests \cite{doran_towards_2010, karlsen_quests_2008, machado_towards_2016}. However, to the best of our knowledge, there have been no attempts to more directly learn how developers think about quests. Talking directly to developers about their views and ideas on quests can provide insight that design patterns cannot. We can ask what a developer is thinking about certain aspects of a quest, instead of making assumptions about their thought process in order to provide a more complete picture of how the industry thinks about quests.

The purpose of this paper is to further understand the definition of a quest in industry settings. Instead of analyzing existing quests for design patterns, we interview developers with experience designing and/or implementing quests or narratives. To the best of our knowledge, this is the first instance of learning about the definition of a quest directly from interviews with developers. We propose that this approach can create better technical definitions for use in industry and academia by incorporating developer feedback. Definitions created in this way may have greater applicability in practical settings, which could increase the adoption of these terms in industry. Cross disciplinary definitions may also help reduce the academic industry gap, and part of this work highlights some gaps in how academics talk abut quests compared to how industry professionals talk about quests. This approach has two main benefits. The first is that we are able to gain a better understanding of how quests are viewed within industry, which has traditionally proved difficult. The second is that we were able to directly ask developers their thoughts surrounding quests, reducing the number of assumptions we have to make about how developers understand quests in comparison to post-hoc analysis. This paper builds off our previous work on a general quest definition by considering the definition from the perspective of developers \cite{yu_towards_2021}. 

%This paper is motivated by recent work on a general quest definition \cite{yu_towards_2021}, and we use our connections in industry to provide insights not available in their work. 

\section{Background}

Video game quest theory was first introduced by \citeauthor{aarseth_hunt_2005} \cite{aarseth_hunt_2005}. This early work was focused on defining a quest, as well as starting to identify design patterns for quests. Initially, the differences between quests in video games and quests in novels were used as the basis for the definition of a quest \cite{aarseth_quest_2004, tronstad_semiotic_2001}. Some researchers brought ideas from other popular media such as Dungeons and Dragons \cite{DnD} or used literary analysis techniques to inform their definition \cite{howard_quests_2008, tosca_quest_2003}. Other areas of early video game quest theory focused on deriving design patterns from RPGs and MMORPGs \cite{doran_towards_2010, smith_situating_2011-1, karlsen_quests_2008}. 

As we mentioned in the introduction, research on quests has lost popularity in recent years. Modern quest theory is done in pieces by many researchers across different areas of interest. For example, quests were analyzed for mechanical or structural similarities in order to more effectively procedurally generate quests \cite{machado_towards_2016, alexander_deriving_2017}. Quest theory has also been used to help design levels \cite{dahlskog_patterns_2015, mourato_challenge_2013} or model player knowledge \cite{lee_effects_2009}. The interest that other research areas have in generalized quest theory highlights the need for academics to revisit this topic. 

Though some researchers who study video game quest theory have considered other perspectives such as incorporating literary analysis into quests, the perspective from industry has not been extensively researched. Developers in industry have important knowledge on the definition of a quest, but in the past it has been difficult for academics to access that knowledge. In most cases, what can be learned from the industry is limited to blog posts, Youtube videos, or the occasional GDC talk \cite{kim_effective_2005}. Many developers do not have time to create these public pieces of information, so their views are not easily accessible. Additionally, this work is rarely focused on general quest theory but instead focused on practical applications and lessons that the developers learned from specific games. This kind of knowledge dissemination is expected due to the fact that industry professionals might be more concerned with how to specifically create quests. This means that the knowledge is less general, and assumptions would have to be made in order to generalize the information. Thus, how developers define quests is largely unknown to the academic community. 

In our previous work, we have developed a technical, general quest definition out of definitions in the academic community \cite{yu_what_2020, yu_towards_2021}. To the best of our knowledge, this is the only quest definition that is specifically defined to support technical applications of quests. Equation 1 and Equation 2 form the basis of this quest definition. Equation 1 defines a quest $Q$ as a set of partially ordered tasks $\leq$ and a distribution of rewards $d(R)$, and equation 2 defines a task $t \in T$ as a set of players $P$ that can complete the quest, the condition $C$ to determine whether a task is complete, the monitoring system $M$ to check the game state, the presentation $I$ of the task, and the distribution of rewards for the task $d(R_t)$. 

\begin{equation}
    Q = \langle T, \leq, d(R) \rangle
\end{equation}

\begin{equation}
    t = \langle P, C, M, I, d(R_t) \rangle
\end{equation}

The early work on quest theory, and many later works done in service for a particular goal, all share similar problems in fully defining a quest. The most common limitation is the context in which the quest definition is situated, or lack thereof. Researchers mostly analyze quests in RPGs or MMORPGs, which limit the applicability of the quest definition to those genres. In the past, it may have been appropriate to limit video game quest definitions to specific games or genres, but in the recent years it has been increasingly common for games to borrow elements from other genres. One of the more popular additions is the inclusion of quests in non-RPG games, such as the Rocket Pass in the sports game Rocket League \cite{RocketLeague}. This illustrates the need to update the definition of a quest to maintain relevance for modern games. Additionally, some quest research ignores context altogether, which we have discovered to be a critical component in understanding quests in industry. If the academic community has a single, generalized quest definition to reference, quests can be more easily applied across research areas such as quest generation, interactive narrative, and player modeling.

\section{Methodology}

We interviewed fifteen developers with industry experience designing and/or implementing quests or narratives in published games. We included narrative development as well as quest development because some companies design narratives using quests. We identified developers to interview in three ways: personal contacts to the authors, through industry contacts of the authors, and through participants in the study. Eight developers were personal contacts of the authors, five developers were provided by industry contacts of the authors, and two were referred from participants in the study. The author who conducted the interviews had no personal connection to any of the developers. Developers were only included in the study if they had professional experience working with quests or narratives in some capacity in at least one published game. We interviewed fifteen developers because that is the number of developers we were able to contact and interview within the time period of this study. 

Each individual was asked the following pre-interview questions about their development experience:

\begin{enumerate}
    \item Do you consider yourself an indie developer?
    \item What is your current role? 
    \item What kinds of design experience do you have? 
    \item How long have you worked in industry? 
    \item Which companies have you worked for? 
\end{enumerate}

We asked these questions because we wanted to know the developer's experience related to quests and narratives. We included these questions to check whether we were able to capture a variety of developer views.

We used a semi-structured interview format. We chose this format because we wanted to allow for follow-up questions and/or clarifications. There could also be some topic that a developer is particularly focused on, and we wanted to allow for the developer to be able to put as much or as little emphasis on particular ideas as they saw fit. During discussion, we used the strategies of asking the developers to more fully explain what they meant by particular terms and repeating their ideas to them, and asking them if they had anything else to add. We asked each developer the following questions:

\begin{enumerate}
    \item In your opinion, what is a quest?
    \item What is the purpose of a quest? 
    \item Do you think that your definition of a quest is the same as other people in the industry?
    \item Does this proposed academic definition of a quest match your definition? In what ways?
    \item Which definition do you prefer?
\end{enumerate}

We asked the developers these questions to try to capture all of the ideas they might have surrounding the concept of a quest. Question 1 was an open ended way to allow the developer to talk about initial thoughts on quests, and discuss the definition of a quest. Question 2 allowed the developer to elaborate on why they think a quest is included in a game. Question 3 asked the developers to comment on whether there is an industry standard quest definition. To further an effort to establish a general quest definition, we asked the developers to compare their quest definition to our recently proposed definition in question 4, and asked which definition they prefer in question 5.

In between questions 3 and 4, we showed each developer a short presentation of our quest definition, which we described in the background section of this paper \cite{yu_towards_2021}. We used our definition as the academic definition because it is a synthesis of several other academic definitions which includes ideas from several researchers.

To process the interviews, we use thematic analysis as described by \citeauthor{braun_using_2006} \cite{braun_using_2006}. To generate codes, two of the authors independently analyzed two interviews to create a starting set of codes. Then, they both analyzed a third interview to identify gaps in the codes. The set of codes was revised, and we calculated interrater reliabilty using a fourth interview. This resulted in a score 0.30, so we further discussed the set of codes for clarification purposes. Finally, we applied this set of codes to 10 excerpts from a fifth interview, achieving an interrater reliability of 0.94. Table \ref{fig:CodeTable} shows the final list of codes that two authors applied to all interviews. 

\begin{table*}[ht]
\centering
\footnotesize
\begin{tabular}{|l|l|}
\hline
Code                                                                             & Description                                                                                                                                                                                                                 \\ \hline
Technical Details                                                                & Discussion of the implementation details within the game to create quests                                                                                                                                                   \\ \hline
\begin{tabular}[c]{@{}l@{}}Background Determines\\ Quest Definition\end{tabular} & Discussion of how their past experiences influence their quest definition                                                                                                                                                   \\ \hline
Goals                                                                            & The types of goals that a player might have                                                                                                                                                                                 \\ \hline
Achievement Goals                                                                 & Goals outside the game but within the platform level                                                                                                                                                                        \\ \hline
Game Goals                                                                       & \begin{tabular}[c]{@{}l@{}}Goals set and communicated to the player within the game, or discussion of the actions that the player takes \\ to complete either a step or the entire goal\end{tabular}                        \\ \hline
Player Goals                                                                     & Discussion of self-set goals by the player                                                                                                                                                                                  \\ \hline
Narrative                                                                        & Any kind of narrative content                                                                                                                                                                                               \\ \hline
Player Experience                                                                & Describing what emotion the player might be experiencing while playing a quest                                                                                                                                              \\ \hline
Player Quest Policy                                                              & What the player does or does not pick, and how                                                                                                                                                                              \\ \hline
Presentation                                                                     & \begin{tabular}[c]{@{}l@{}}How the quest is presented to the player, and what information is presented to the player\end{tabular}                                                                                        \\ \hline
Gameplay Elements                                                                & \begin{tabular}[c]{@{}l@{}}Discussion of specific gameplay elements that can be used to present the quest to the  player, or discussion of \\ specific in-game elements that can be used as a part of the quest\end{tabular} \\ \hline
Information Transfer                                                             & \begin{tabular}[c]{@{}l@{}}Discussion of the knowledge needed to complete the quest, or knowledge transferred using the quest\end{tabular}                                                           \\ \hline
Progression                                                                      & Game or player progression                                                                                                                                                                                                  \\ \hline
Purposes                                                                         & What is the quest for                                                                                                                                                                                                       \\ \hline
\begin{tabular}[c]{@{}l@{}}Quest Definition\\ Varies By Use Case\end{tabular}    & Quest definition changes based on the design needs, the game, conversation, or context                                                                                                                                      \\ \hline
Quest Types                                                                      & Any way of categorizing types of quests                                                                                                                                                                                     \\ \hline
Critical Thinking                                                                & \begin{tabular}[c]{@{}l@{}}Thought process in the creation of their quest definition, or creating their definition during the interview\end{tabular}                                                                      \\ \hline
Reward                                                                           & The concept of reward                                                                                                                                                                                                       \\ \hline
Extrinsic Reward                                                                 & What the player gets in the game                                                                                                                                                                                            \\ \hline
Platform Reward                                                                  & Something outside the game, but still external to the player                                                                                                                                                                \\ \hline
Intrinsic Reward                                                                 & What emotions/satisfaction does the player feel from doing something in the game                                                                                                                                            \\ \hline
\end{tabular}
\caption{Table of the codes that were used to analyze the interviews}
\label{fig:CodeTable}
\end{table*}

\section{Interview Analysis}

We analyze and discuss the answers to the pre-interview questions and the first three interview questions here. 

\subsection{Pre-Interview Questions Analysis}
The developers that we interviewed represent a variety of experience. Six developers identify as indie developers, seven developers do not consider themselves indie developers, and two developers have experience with both. The reported minimum amount of industry experience is three years, and the maximum is over 30 years. Three developers are experienced in systems and programming, one developer has both programming and design experience, two developers are in a production role, and nine developers have experience designing quests and narratives. These developers have worked at studios including Bioware, Bungie, and Ubisoft, as well as eight different independent studios. 

We assign each of the fifteen developers a label based off of their pre-interview questions. The label has the format ``$<$Type of Industry Experience$><$Number of years experience$>$", where we use the prefix ``IND" to denote developers who consider themselves indie developers, ``AAA" to denote developers who do not consider themselves indie developers, and ``BOTH" for both. Additionally, in the case where the type of industry experience and the number of years of experience is the same between two developers, we append ``.1" and ``.2" to the label to distinguish the two.

\subsection{Quest Definitions}

We asked each developer what they think a quest is in order to collect initial thoughts on quests and discuss the definition of a quest. We considered the answer to this question to be each developer's personal definition of a quest. This allows us to determine if there is a consensus on the definition of a quest within the developers we interviewed. To help answer this, we use the following labeling system which identifies elements in a quest definition \cite{yu_what_2020}. We used this labeling system because it was the same labeling system that we previously used to identify elements in academic definitions. 

\begin{itemize}
  \item T - Tasks, objectives, goals
  \item R - Rewards and progression
  \item P - Player and player experience
  \item N - Narrative 
  \item D - Designer or developer
  \item O - Other
\end{itemize}

We added to our original labeling system in order to better capture the aspects the developers talked about in their quest definitions. We extend the letter ``P" to include the notion of player experience, and add a new label ``D" which denotes if the definition explicitly refers to a designer or developer. For example, IND08 specifically used the word ``designer" in their definition. We apply a label to a quest definition if it contains that element. Table~\ref{fig:QuestDefinitionTable} gives the labels applied to each quest definition. 

\begin{table}[ht]
\footnotesize
\centering
\begin{tabular}{|l|l|l|l|l|l|l|}
\hline
Developer & T & R & P & N & D & O \\ \hline
IND04     & X & X & X & X &   &   \\ \hline
IND05     & X & X & X &   &   &   \\ \hline
IND06     & X &   & X &   &   &   \\ \hline
IND07     & X &   &   &   &   &   \\ \hline
IND08     & X & X & X &   & X &   \\ \hline
IND27     &   &   &   &   &   & X \\ \hline
BOTH07    & X &   & X &   & X &   \\ \hline
BOTH08    & X & X & X & X &   &   \\ \hline
AAA03     & X & X & X & X &   &   \\ \hline
AAA14.1   & X &   & X &   &   &   \\ \hline
AAA14.2   &   &   &   &   &   & X \\ \hline
AAA16.1   &   &   & X & X &   &   \\ \hline
AAA16.2   & X &   & X & X &   &   \\ \hline
AAA19     & X & X & X &   &   &   \\ \hline
AAA30     & X & X & X &   & X &   \\ \hline
\end{tabular}
\caption{Application of the labeling system to each quest definition}
\label{fig:QuestDefinitionTable}
\end{table}

There are two definitions that are part of the other category because they do not include any of the elements in the labeling system. The quest definition provided by IND27 is ``anything that happens between `A' and `B'", and the quest definition provided by AAA14.2 is ``The reason to play a video game." Even though these definitions are both different from the other definitions provided by the developers, the two developers themselves do not share similarities in work experience. Additionally, there are no distinct patterns or grouping in the quest definitions given this labeling system.  

\begin{figure}[ht]
    \centering
    \includegraphics[width=0.5\textwidth]{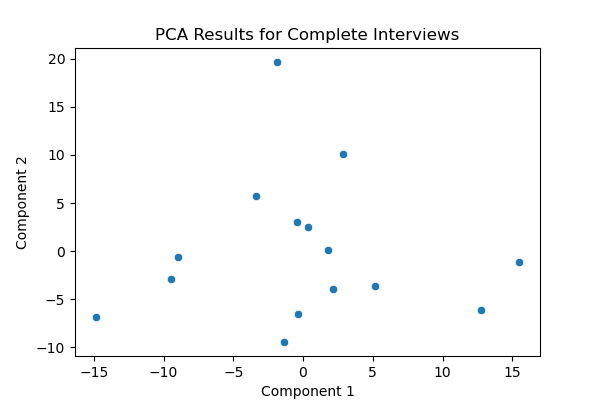}
    \caption{PCA results on the code vectors of each interview}
    \label{fig:PCA}
\end{figure}

To further understand the variability of the quest definitions, we use principle component analysis (PCA) on the interviews as a whole, shown in Figure \ref{fig:PCA}. We treat each interview as a vector, where the index of the vector is a given code, and the value is the number of excerpts with that code. These vectors provide some idea of what developers consider important based on how frequently they discussed a particular topic.

We analyze each component to understand which codes have a high impact on differentiating vectors along that component, and identify a value of 0.25 or greater to be a significant contribution. Component 1 is most affected by the codes ``Critical Thinking", ``Game Goals", ``Player Experience, and ``Purposes". Component 2 is most affected by ``Progression", ``Narrative", and ``Extrinsic Reward". There are a few interviews that may be close together, but there are no distinct groups or clusters in these results. 
The variability in the quest definitions, and the variance in the PCA results both support the idea that there is significant variety in the beliefs of the developers. Generally, this signals that there is no consensus on a quest definition in the industry. The type of industry environment or years of experience does not correlate to any trends. We found that designers have thought more deeply about how quests interact with design than the definition of a quest. A common type of statement in the ``Critical Thinking" excerpts was ``I hadn't really thought about [quests] in this way ... trying to define [quests] very rigorously before" (AAA03). On the other hand, many developers were able to confidently express how to design what they considered a ``good" quest. 

This conclusion is further supported by the seven developers who created or modified their definition during the interview, commenting things like ``here's something that was missing from my definition earlier..." (BOTH08). These represent half of the interviews, which indicates that a significant portion of developers needed to adjust their quest definition when asked to think critically about quests.

Despite the variety and lack of a single, cohesive quest definition, Table \ref{fig:QuestDefinitionTable} shows some overlap in ideas. Tasks, rewards, and players each appear in half or more of the definitions provided by the developers. This suggests that a general quest definition is possible, and these common ideas should be included in the definition. 

\subsection{Expected Definition Similarity}

We asked every developer whether they thought their quest definition matched other individuals in industry. Seven developers expected that their definitions would not match, five developers expected that their definitions would be different but with overlap, and two developers expected that their definition would match. One developer said they did not know. % A majority of developers believed that there are differences in quest definitions, with half believing there is no overlap at all, and a third believing that there is some overlap. 

The majority of developers, twelve, also stated that there is no singular quest definition within industry. Some of the developers provided some reasoning as to why that might be the case. Four developers posited that quest definitions differ due to the experience of the developer, and two propose that the difference is due to the game itself. AAA03 states ``I strongly suspect that people have some different elements of their definition based on what games they've played or which games they first encountered", and IND05 comments ``my definition of a quest depends on what I believe a game is to begin with, then inherently everyone's going to have a different definition."

\subsection{Purpose of a Quest}

We asked each developer about the purpose of a quest. Developers often provided multiple purposes, and the most popular ones are shown in Table \ref{fig:PurposesTable}. Some other proposed purposes of a quest included ``Immerse the player in the game" and ``Provide structure to the game", but these were only mentioned once.

\begin{table}[hbt!]
\centering
\footnotesize
\begin{tabular}{|l|c|}
\hline
Purpose of a Quest                                                                         & \multicolumn{1}{l|}{Number of Interviews} \\ \hline
\begin{tabular}[c]{@{}l@{}}Guide the player through the game\end{tabular}               & 5                                         \\ \hline
Tell the narrative/story                                                                   & 4                                         \\ \hline
\begin{tabular}[c]{@{}l@{}}Provide content to engage the player\end{tabular}   & 4                                         \\ \hline
\begin{tabular}[c]{@{}l@{}}Teach the player rules of the game\end{tabular}          & 4                                         \\ \hline
\begin{tabular}[c]{@{}l@{}}Provide narrative context for \\ gameplay\end{tabular}          & 4                                         \\ \hline
\begin{tabular}[c]{@{}l@{}}Provide reason to play the game\end{tabular} & 2                                         \\ \hline
Progress through the game                                                                  & 2                                         \\ \hline
\end{tabular}
\caption{Proposed purposes of a quest}
\label{fig:PurposesTable}
\end{table}

\section{The Definition-Context-Purpose Paradigm}

\begin{figure}[b]
    \centering
    \includegraphics[width=0.45\textwidth]{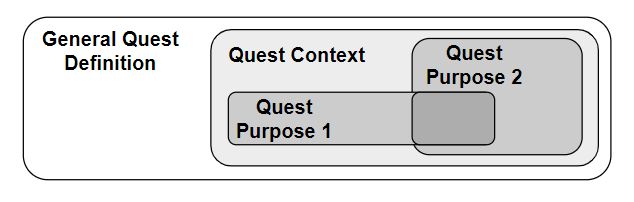}
    \caption{Illustration of the sets defined by the definition-context-purpose paradigm}
    \label{fig:DCP}
\end{figure}

From results of the first part of the interview we identify the three major themes of quest definition, quest context, and quest purpose. These themes provide insight into how industry professionals think about quests. We organize these themes as the definition-context-purpose paradigm for understanding quest design in video games, shown in Figure \ref{fig:DCP}. This paradigm is not intended to be used as a guide for how to specifically design a single quest, but rather a tool to understand broad trends or patterns that may appear in quests across games. To explain this paradigm, we first provide an overview and example here. Then, we discuss each part of the paradigm individually, and end with a discussion of some of the relationships between these themes. 

The first part of the paradigm, definition, acknowledges the existence of a general quest definition which can be applied to games across genres. This general quest definition is a collection of elements that defines a set of all objects that could be considered a quest in video games. The context part of the paradigm narrows the general set of quests into a subset of quests that are appropriate for a particular game. A game can have multiple contexts, such as the story missions and bounties in Destiny 2 \cite{Destiny2}. Third, the purpose part of the paradigm further specifies the quest context, where the purposes define the subset within a given context that have that purpose. A single context can have multiple purposes in it, and these purposes can overlap. The breakdown of the subsets of quests that this paradigm can represent is shown in Figure \ref{fig:DCP}.

Let us imagine we want to design quests for a science fiction RPG. For definition, we will use a general quest definition formed out of the popular common elements provided by designers, as shown in Table \ref{fig:QuestDefinitionTable}. We define a quest to be any game object that includes a task and a reward, and is intended for a player. For context, we are using a science fiction setting so we want our quests to be referred to as missions. We can add more constraints to the context such that every mission will include some narrative, will contain at least three tasks, and the reward will be experience points. For purpose, we use the following: teach the player something, immerse the player in the world, and provide a reason to play the game. Now that the three parts of the paradigm have been decided, we can design quests. The first mission in this game is the tutorial mission, so we assign the purpose of ``teach the player something". We assign later missions the purposes ``immerse the player in the world" and ``provide a reason to play the game." 

\subsection{Quest Definition}

The first major theme is the idea of a general quest definition as a collection of required elements. Half of developers believe that there is overlap between the definition that they provided, and other definitions in the industry. AAA16.1 believes that ``we would all mostly agree what a quest is", but the precise amount of overlap or agreement is unknown to the developer. IND27 agrees, stating that ``it would be fairly different between different people, but I'd think that there would be a lot of similarities''. 
This concept of a general quest definition aligns with our previous research in the area \cite{yu_what_2020, yu_towards_2021}. 

Despite the variety and lack of consensus, there is some commonality in the quest definitions. Table \ref{fig:QuestDefinitionTable} supports this idea, showing a high degree of agreement on the elements task, reward, and player. Twelve developers included some task, objective or goal, seven included a reward, and twelve included the idea of a player. These elements represent significant overlap between ideas, despite the individual definitions and interviews lacking distinct trends. 

\subsection{Quest Context}

The second major theme is quest context, which refers to the specific constraints designers enforce on quests for a particular game. AAA16.1 discusses this explicitly, saying ``right now my company [is] defining what are our quests. What are the structures? What must a quest designer do to make a successful quest?" AAA14.1 also agrees, stating that ``how [quest definitions] differentiate is what kind of game we're kind of trying to make. And what people decide makes sense for their game." This idea was captured by the code ``Quest Definition Varies by Use Case", which appeared 57 times across thirteen interviews. 

The constraints for the quest context can include things like whether the quest should have narrative or what kinds of presentation the quest must have, and often is used to provide guidelines for how to design a ``good" quest.  BOTH08 describes their personal context for good quests, saying a quest ``doesn't even need to progress the story, but [they] like the ones where [quests] do." The codes ``Game Goals", ``Narrative", and ``Information Transfer" have the highest co-ocurrence with the ``Quest Definition Varies by Use Case" code, with 18 times, 10 times, and 9 times respectively. These represent some of the constraints that can be considered context for a quest. For example, a narrative constraint for the quest could be that the quest needs to tell some of the main story. The quest context also includes the practical limitations of creating quests for a game that will be published. For example, the context of a quest could be the limitation that the quest cannot contain a cutscene or voice acting, as they are very expensive to produce.

%The context of a quest is used to narrow the set of quests from the general quest definition into a subset of quests that are usable for a game. 

The quest context represents a divergence in understanding between academia and industry. In academia, consideration for the context of the quest is often ill-defined. For example, quests can be procedurally generated using narrative cohesiveness as a focus \cite{BreaultOuelletDavies2021}, but do not provide any other context. There are many other factors that developers consider important context. For example IND08's entire experience is in designing quests for single player games, which they consider important context as they ``have no concept of how to design a multiplayer game." This suggests that procedural quest generation research which focuses on specific, well-defined contexts might be more accessible to industry than attempts to create a general purpose quest generator.

\subsection{Quest Purpose}

The last major theme is the purpose of a quest, which is the reason or reasons that the quest is in the game. The purpose or purposes of a quest describes how the quest functions within the greater game design space and helps justify its inclusion to the game.For example, lets assume that the first quest at the beginning of a game has two purposes: to teach the player a game mechanic, and to tell the player the premise of the game. These two purposes indicate which part of the player experience they are addressing, as well as indicating which design problems the quest is meant to address. The purpose ``To teach the player a game mechanic" addresses the problem of ``how does the player know how the mechanics work", and the purpose ``to tell the player the premise of the game" addresses the problem of ``how does the player understand the story".  AAA14.1 describes the importance of purpose to the design space by asking ``what would I prefer to make a good quest? Probably more like, what's the purpose?". This shows that the purpose is the most useful part of the paradigm to help design the quest object itself. The quest design problem is equivalent for many of the developers in this study to how the quest fulfills a particular purpose or purposes.

In general, developers intuit a structure where a single quest object might have multiple purposes, and the purposes are as important as the quest itself. Therefore, defining a quest as ``as a guide to help give players direction" (IND06) provides insight to the developer's understanding of quests despite the fact that the definition does little to inform us about the elements of the quest. This also alludes to the idea that the actual quest object itself might not be as important as the reason the quest is being designed. IND06 outright stated that purposes are more important than the elements of a quest, saying ``I don't think about what they are so much as what they're for." Thus, the focus on the purpose of a quest represents a stark divergence between academia and industry.

\subsection{Relationships Between Themes}

IND04 describes the relationship between the definition and context by noting ``if we talked to the game designers on my team they would have a very, very different definition of the word quest because I'm thinking of it way more generally. It's like if I was talking about a system, like the most abstract version of a quest. I go talk to a designer and he [says] `no, quest is literally this, you talk to this character in the game, they tell you to go do a thing. You go do a thing, you come back to them and that is a quest'". IND04 specifies that they are thinking about quests in the definition space, which is the most general way to approach a quest. The hypothetical designer they are referring to is instead considering quests within a specific context, as indicated by the constraints this designer enforces on quests. 

AAA14.1 describes the relationship between context and purpose by saying ``I don't want to define a quest, like it's a list of things in your quest log...because to me it's not about the implementation". They then discuss overcoming their own bias of wanting to initially define a quest solely by context, then elaborate that they would rather consider the question ``what do you want the player to experience?" AAA14.1's comments indicate that the context alone is not enough to fully understand quests in video games. They initially discuss the quest context, and reject the notion that the context of the quest is enough to design quests. They then bring up the purpose of a quest by posing question about player experience. We intend the ``purpose'' part of our framework to address AAA14.1's question about player experience.

\section{Academic Quest Definition Comparison}

The definition-context-purpose paradigm relies on the existence of a general quest definition, and we propose the academic quest definition from section 2 as a possible definition. We gave each developer a short presentation to explain this definition, which included a description of the elements and an example. In this section, we discuss the responses to the final two questions in the interview. 

\subsection{Academic Quest Definition Response}

We asked each developer if they considered their definition to match the academic definition. We show the responses in Table \ref{fig:QuestResponseTable}. None of the developers explicitly stated that the academic quest definition was completely divergent from their own beliefs. The four developers that did not comment on the amount of overlap between definitions were instead focused on testing the limits of the academic quest definition, and the interviewer did not have the opportunity to refocus the conversation.

\begin{table}[b]
\centering
\footnotesize
\begin{tabular}{|l|c|}
\hline
Quest Definition Response & \multicolumn{1}{l|}{Number of Interviews} \\ \hline
Overlap in Definitions    & 6                                         \\ \hline
Definitions Match         & 5                                         \\ \hline
Broader Than Their Definition       & 4                                         \\ \hline
\end{tabular}
\caption{Response to whether the academic quest definition matches the developer's definition}
\label{fig:QuestResponseTable}
\end{table}

 % on whether the developer considered their definition matched the academic definition. 
 
Four developers mentioned that the academic quest definition was broader than their own definition. The academic quest definition is intended to be applicable across genres, so this offers some support that the definition is general. However, five developers attempted to determine the limits of its applicability, which suggests that the definition could become more general in future iterations.

Of the developers who considered their definition to match, one developer provided a definition that is substantially different. Every other developer provided definitions which specified elements of a quest, but AAA30 stated a quest is ``a diegetic way to structure the player experience". This definition does not include any description of the elements that are present in the academic quest definition, and instead is a description of the purpose of a quest with the context of diegetic elements. If we assume that the developer is strictly speaking about the definition of a quest, then the developer's perception that the two definitions match may seem incorrect. However, using the definition-context-purpose paradigm, we can interpret this conflict as the developer focusing on the purpose part of the paradigm, which they felt fit into the definition part of the paradigm. 

\subsection{Quest Definition Preference}

The last question we asked developers was which quest definition they prefer. Unfortunately, we did not ask all of the developers this question due to time constraints. Eight developers answered this question. One developer preferred the academic definition, and one developer preferred their own definition. Six developers said that their preference depended on the use case of the definition. For example, AAA19 said that they "like [the academic quest definition] for taking a step back and just thinking about all the parts of the quest. But then [they] would break it down into less academic language if [they were] actually going to get someone to implement a quest." This suggests that despite the strengths of the academic quest definition, it may be too general for a specific quest implementation. Given the definition-context-purpose paradigm, this finding is unsurprising because the context plays a major role in producing usable quests.  

\section{Conclusion}

In this paper we interviewed fifteen developers with quest or narrative experience to understand the definition of a quest in industry. These developers provided important insight into the definition of a quest that has been previously unknown to the academic community. There was high variability in the topics discussed by the developers, but, despite this, we identified three major themes surrounding quests and introduced these themes as the definition-context-purpose paradigm. This paradigm proposes how the these trends fit together to explain how developers understand a quest. Finally, we discussed how the developers generally reacted positively to a proposed general quest definition, where a majority of developers viewed the definition as equivalent or better than their own. In future work, we would like to take the lessons learned from these interviews and incorporate them into the academic quest definition to increase its generality and usability in industry settings.

\bibliography{Main.bib}

\end{document}